\documentclass[doublecol,figures]{epl2} 

\title{Ribbon polymers in poor solvents:  layering transitions in annular and tubular condensates}
\shorttitle{} 

\author{Y. Y. Suzuki\inst{1,2} \and D. R. M. Williams\inst{3}}
\shortauthor{Y. Y. Suzuki \etal}

\institute{                    
  \inst{1} Institut de Physique Th\'{e}orique, CEA - IPhT, CNRS, URA 2306  F-91191 Gif-sur-Yvette, France\\
  \inst{2} Faculty of Engineering, Takushoku University - Hachioji, Tokyo 193-0985, Japan\\
  \inst{3} Research School of Physical Sciences, Australian National University - Canberra, ACT 0200, Australia\\
}

\abstract{
We study the structures of a ribbon or ladder polymer immersed in poor solvents.  
The anisotropic bending rigidity coupled with the surface
tension leads ribbon polymers to spontaneous formation of highly anisotropic condensates
in poor solvents.
Unlike ordinary flexible polymers these condensates undergo a number of
distinct layering transitions as a function of chain length or solvent
quality, and the size of condensates becomes non-monotonic function of
chain length.  We show that the fluctuations of the condensates are in
general small and these condensates are stable.}

\begin{document}

\maketitle

Recently considerable attention has been devoted to the physics of
semiflexible polymers, {\it i.e.}, chains which have significant bending
stiffness. These kinds of polymers exhibit interesting physics in the
form of liquid crystalline phases and are of great importance in
biology and biophysics.  For example, DNA is the most well-known
semiflexible polymer.  
Much of the interest has focused on
semiflexible chains with isotropic bending elasticity\cite{toroid0,toroid1,Frank-Kam,MarkoSiggia,Odijk0,Marko,DNAbend}. Macroscopically
this corresponds to modeling the chain as a cylinder with a circular
cross section.  In many cases, however, the cross section is highly
anisotropic\cite{Exp,Nyrk1,Nyrk2,Nyrk3,Goles}. 

\begin{figure}
\onefigure[width=7cm]{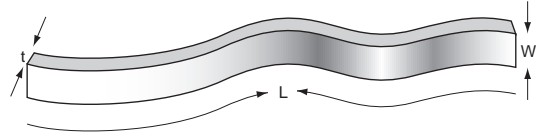}
\caption{Model for anisotropic semiflexible polymer}
\label{ribbon}
\end{figure}

The anisotropic polymer can be modeled as a ribbon, of length $L$, width $w$,
and thickness $t$ assuming $L \gg w > t$ (fig.\ref{ribbon}) 
with anisotropic bending elasticity.  Indeed, if the
ribbon is composed of an isotropic material, the ratio of the elastic
constants for bend in the easy and hard directions is
$\epsilon_e/\epsilon_h = (t/w)^2$ \cite{Landau} . This strong
dependence on the ratio of $t/w$ implies that the ribbon polymer shows
a very anisotropic elastic response. On the chemical scale, the chains
do not normally consist of an isotropic material, and details of
chemical bonding are important for bend, thus, even higher ratios of
$\epsilon_e/\epsilon_h$ are possible.

Some of the solution properties, in particular the
liquid-crystalline behavior, have been studied theoretically
\cite{Nyrk2}.  Here we study the structure of a single ribbon polymer
immersed in a poor solvent.  Assuming the system is well below the $\Theta$
temperature, the chain collapses completely and forms a condensate which
is effectively a polymer melt.

For isotropic semiflexible chains such as DNA, the condensate often forms a toroid
\cite{toroid0,toroid1,toroid2,toroid3,Li,Ubbink,Hud,Park,toroid4,Takenaka} 
and sometimes a globule\cite{globule1,globule2,Sikorav,Odijk04} 
through intermediate states\cite{intermed1,intermed2,intermed3}.  
For ribbon polymers, we show
in this paper that the condensates forms an annulus. The annulus
formation manages to avoid bend in the hard direction, while reducing
contact of the chain with the solvent.  The novel character of these
annular condensates is that multilayer annuli can form and they
undergo sudden changes as a function of chain length or surface
tension. This is in marked contrast to the case of ordinary flexible
or semiflexible polymers where no dramatic changes occur.  This
quantization is a direct consequence of the anisotropic elasticity.

To model the system, we introduce two surface tensions $\gamma_h$ and
$\gamma_e$ which corresponds to the surface energies of contact between
the top (narrow surface) of the polymer and the solvent, and the
surface energies of contact between the side (flat and wide surface)
of the polymer and the solvent, respectively.   If the chain consists of anisotropic
molecules, any ratio of $\gamma_e / \gamma_h$ is possible in principle.
In this paper, however, we assume $\gamma_e / \gamma_h \approx 1$.

A large number of parameters appear in this model.  Once final results have been obtained, it is useful to ignore
numerical prefactors and to make some crude estimates. 
Then, we write $w
\approx t \approx a$ and $\gamma_e \approx \gamma_h \approx g kT/a^2$, where $g$ is of order
 unity and $a$ is a length of order of one
Angstrom. 
It is convenient to define two bare persistence lengths,
$l_e \equiv \epsilon_e/kT$ and $l_h \equiv \epsilon_h/kT$ which are
essentially the scales on which a free chain bends in the two
directions due to thermal fluctuations.

We begin by examining the shape of condensates when we slowly lengthen
the chain. For short chains, bend costs elastic energy, then the chain
more-or-less remains as a rod. For slightly longer chains, however,
the overlap induced by bend compensates the elastic energy by reducing
its surface energy, therefore, the chain may form a ring. For a circular
ring of radius $R$ it is easy to show that the bending energy is ${ 1
\over 2} \epsilon_e L/R^2$. The change in surface energy in bending
into a ring is $4 \pi R w \gamma_e - 2 L w \gamma_e$.  
 Minimising it over $R$ and equating the free energies  
gives the critical length, $L^*= (3 \sqrt{3} \pi / 2) \sqrt{ \epsilon_e/(\gamma_e w)}$  
at which ring formation occurs. The chain overlaps by $1/2$ turn, at
$L=L^*$. Our interest here is in the case $L \gg L^*$ so the chain
winds around many times.

\begin{figure}
\onefigure[width=6cm]{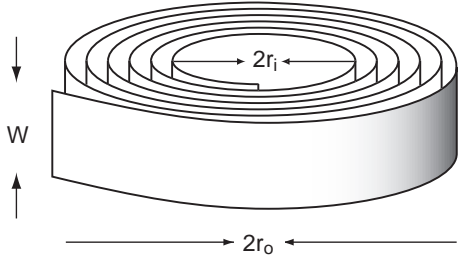}
\caption{Annular condensate for short ribbon polymer in poor solvent}
\label{disk}
\end{figure}

The simplest way the chain can pack is to form a disk with a hole in
the middle (annulus) as shown in fig.\ref{disk}. This annulus has
height $w$ and inner and outer radii $R_i$ and $R_o$,
respectively. The total energy consists of three terms: (1) surface
energy of the top and bottom surfaces, (2) surface energy of the
inner and outer exposed sections of the annulus, (3) bending
energy of the chain.  This last term is due to bending only in the
easy direction. 

Because the volume of the chain is conserved, $R_o$ and
$R_i$ are related by
\begin{equation}
t L w = w \pi (R_o^2 - R_i^2).
\end{equation}
There is thus only one variable to minimize over (say $R_i$). Also,
since the area of the upper and lower surfaces is fixed ({\it i.e.}, both are
always of area $tL$), there are really only two relevant terms in the
total energy. The lateral surface energy is easily calculated as
\begin{equation}
2 \pi \gamma_e w (R_o+R_i).
\end{equation}
The bending energy is calculated by
\begin{equation}
 \frac{ \epsilon_e}{2} \int_0^L ds \left( \frac{d \theta}{ds} \right)^2,
\end{equation}
where $s$ is the arc length and $\theta$ is the angle
between the tangent of the ribbon curve and some fixed direction in the
plane of the disk. 

In one complete turn of the ribbon, $\theta$
increases by $2 \pi$ and $s$ increases by $2 \pi r$ where $r$ is the
distance of the chain from the center of the annulus. This gives $
d\theta/ds = 1/r$. During the same change in $s$ the radius increases
by one ribbon thickness, {\it i.e.}, $dr/ds= t/(2 \pi r)$.  This allows us to
write the bending energy as $ \pi \epsilon_e t^{-1} \ln(
R_o/R_i)$.
The total energy is then
\begin{equation}
U = \frac{\pi \epsilon_e}{ t} \ln \left( {R_o \over R_i} \right)+ 2 \pi \gamma_e w
(R_o+R_i),
\end{equation}
subject to $R_o^2=  R_i^2+ t L/\pi$. 

It is convenient at this point to introduce two characteristic
lengths. One is $\Lambda \equiv \epsilon_e/(2\gamma_e w t)$, and the other
is $\Gamma \equiv \sqrt{Lt/\pi}$. 
For long chains, $\Gamma$ is much larger than $\Lambda$, and we shall see $R_i \approx \Lambda$ and $R_o \approx \Gamma$.  
In this limit, $\Lambda \ll \Gamma$, the annulus is ``fat'', {\it i.e.}, it has a small hole
in the middle. In the opposite limit, $\Lambda \gg \Gamma$, the hole is large and the annulus
is ``thin''.  It looks
like a ring.

In the long-chain limit, $R_i \ll \Gamma$, the total energy is approximated by
\begin{equation}
U= {\rm const.} + 2 \pi \gamma_e w R_i - \frac{\pi \epsilon_e}{ t}
\ln \left( {\pi R_i \over \sqrt{Lt}} \right),
\end{equation}
which has a minimum at $ R_{i}^* =
\Lambda$. Using our crude estimates give $ R_i^* \approx l_e/g$,
{\it i.e.}, the internal radius is somewhat less than the bare persistence
length. In this limit the internal radius of the annulus is
independent of chain length, whereas the external radius is
approximately $R_o \approx (tL/\pi)^{1/2}$. A more accurate expression
for $R_i^*$ can be found by expanding in ascending powers of $\alpha
\equiv \Lambda/\Gamma $.  This gives $R_i^* = \Lambda( 1 - \alpha + {
3 \over 2} \alpha^3 - ...)$.  From this it is clear that, even though
$R_i^*$ becomes asymptotically $\Lambda$, the approach is rather slow
as $L^{-1/2}$.

In the limit $ \Lambda \gg \Gamma$, we can again expand the total energy
and find the minimum at $R_i^* = (\Gamma^2 \Lambda/2)^{1/3}$, or more
crudely, $R_i^* \approx ( L l_e a/g)^{1/3}$.  The crossover between
the thin and fat regimes occurs at $\Gamma=\Lambda$. A crude estimate
of this gives $L_{c} = a (L^*/a)^4$. This implies that the chain
must be rather long to be in the fat regime.

\begin{figure}
\onefigure[width=4cm]{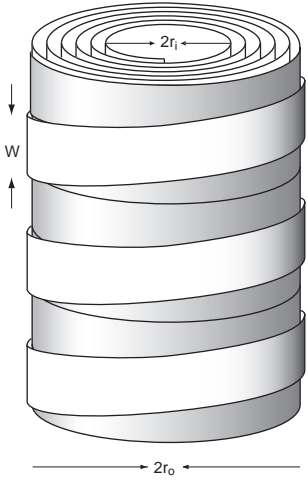}
\caption{tubular condensate for long ribbon polymer in poor solvent}
\label{tube}
\end{figure}

The results just derived correspond to the case of one layer. The
novel point about this system is that it can choose to form more than
one layer. This decreases the surface area exposed by the top and
bottom surfaces but comes at the expense of extra surface energy on
the inside and outside surfaces and extra bend both in the hard and
easy directions.  For second layer to form the chain must bend in the
hard direction. The best way of doing this is to join the two layers
together at the outer radius (fig.\ref{tube}).  

We need to calculate the bending energy
of the chain in the hard direction. The chain must bend between the
two layers. It must jump a distance $w$ in the vertical direction
while winding a distance $\Omega = 2 \pi R_o$ in arc length.  If we
let $\phi$ be the angle made by the chain with the plane of the lower
annulus then the bending energy can be written as 
\begin{equation}
 U_{h} =  \frac{\epsilon_h}{2} \int_0^\Omega ds \left( {d \phi \over ds} \right)^2.
\end{equation}
This is to be minimized subject to a constraint that the chain jumps up a distance
$w$, {\it i.e.},  $ \int_0^\Omega ds \phi = w$.  Here we have made the very
good approximation that the angle $\phi$ is always small. Using the
method of Lagrange multipliers and the Euler-Lagrange equations we
find that the minimum energy trajectory coresponds to the
solution of the differential equation: $d^2 \phi/ds^2 = {\rm
const}$.  Using the boundary conditions $\phi(0)=0=\phi(\Omega)$
the constraint gives 
\begin{equation}
\phi(s) = \frac{6 w}{\Omega} \left\{ \frac{s}{\Omega} - \left(\frac{s}{\Omega} \right)^2 \right\},
\end{equation}
and the bending energy becomes
\begin{equation}
U_{h} = {6 \epsilon_h w^2 \over \Omega^3}. 
\end{equation}
It is fairly clear from this formula that the chain
wants to spread the bend over as much length as possible, so that it
is best to bend at the outer radius of the annulus and $\Omega = 2 \pi
R_o$.  

We now examine the transition from one to two layers. The
one-layer energy is
\begin{equation}
U_1 = {\pi \epsilon_e \over t} \ln \left({ R_o \over R_i} \right)+ 2 \pi \gamma_e w (R_o+R_i)
+ 2 \gamma_h L t,
\end{equation}
subject to $R_o^2=  R_i^2+ t L/\pi $.
The two-layer energy is
\begin{eqnarray}
U_2 &=&  {2 \pi \epsilon_e \over t } \ln \left( {R_o \over R_i} \right)+ 4 \pi \gamma_e w (R_o+R_i) \nonumber \\ 
&&+  \gamma_h L t +{6 \epsilon_h w^2 \over (2 \pi R_o)^3}  ,
\end{eqnarray}
subject to $R_o^2 =  R_i^2+  t L/2\pi $.
Here $\gamma_h L t$ is the surface tension of the ``top'' of annulus.

The transition from one to two layers occurs when $U_1 =
U_2$.  To solve this equation we need to minimize $U_1$ over $R_i$ and
similarly  for $U_2$. In principle different $R_i$ will be associated with
the two geometries. However, in the limit where $R_i \ll (Lt)^{1/2}$
the result is the same for both, {\it i.e.}, $R_i^2 \approx \epsilon_e/(2
\gamma_e w t)$.  This gives
\begin{eqnarray}
U[n] &=& \frac{n \pi \epsilon_e} {t} \left\{ 1 + \ln\left( {2 \gamma_e t w \over \epsilon_e } \sqrt{ L t \over (n \pi)}\right) \right\} \nonumber \\
&&+ \frac{2}{n} \gamma_h L t + 2 \gamma_e \pi w \sqrt{ n L t \over \pi} \nonumber \\
&&+ {3 (n-1) \epsilon_h w^2 \over  \sqrt{2} \pi^3}  \left( Lt \over \pi \right)^{-3/2} ,
\end{eqnarray}
where $n$ is the number of layers ($1$ or $2$). 
Equating the two energies for $n=1$ and $n=2$, and ignoring the logarithmic factors,
gives
\begin{eqnarray}
 {\pi \epsilon_e \over t} - \gamma_h L t + 2 ( \sqrt{2} -1) \pi w \gamma_e \sqrt{Lt \over \pi} && \nonumber \\
 + {3 \epsilon_h w^2 \over \sqrt{2} \pi^3} \left( Lt \over \pi \right)^{-3/2} = && 0 .
\end{eqnarray}
The four terms in this equation represent the four kinds of energy
differences between the two layer and one layer system. In order these
are: Easy bend; top/bottom surfaces; side surfaces; hard bend.  The
driving force for the transition to two layers is the top/bottom surface
energy.  This is opposed by the other three terms.  We are mainly
interested here in the case where the hard bend is the dominant
penalty. In this case the transition occurs when
\begin{equation}
L^5 \approx {\epsilon_h^2 w^4 \over (\gamma_h^2 t^5)},
\end{equation}
where we have
ignored numerical prefactors.  Even more approximately this can be
written $ L \approx a (l_h/a)^{2/5}$ where $l_h$ is the persistence
length in the hard direction. Provided $l_h \gg l_e$ (it is true when
$L \gg R_i$) our approximations are valid.

At the transition the jump in $R_i$ is small, since $R_i $ is about
the same for both layers. However, there is a sudden jump in $R_o$ by
a factor of $1/\sqrt{2}$.  In the case of a three layer system, the
chain must jump once on the outside radius and once on the inside
radius. Since the jump on the inside radius must take place over a
shorter amount of arc length, its energy cost is expensive. 
The energy for $n$ layers is then
\begin{eqnarray}
U[n] &=& {n \pi \epsilon_e \over t} \ln \left( R_o \over R_i \right)  \nonumber \\
&& + n (2 \pi \gamma_e w) (R_o+R_i) + { 2 \over n} \gamma_h L t \nonumber \\
&&+  \frac{6 \epsilon_h w^2}{ (2 \pi)^{3}} \left( 
\left[ {n \over 2}\right] R_o^{-3} + \left[{{n-1} \over 2}\right] R_i^{-3}
\right),
\label{nlayer}
\end{eqnarray}
with
$ R_o= ( R_i^2 + n^{-1} tL/\pi)^{1/2}$,
and where $[x]$ indicates the integer part of $x$. 

\begin{figure}
\onefigure[width=6cm]{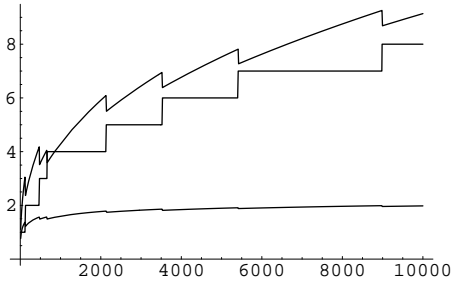}
\caption{Layering transition in a fat condition $\Gamma/\Lambda \approx 10$.  The horizontal axis is the polymer length $L/w$.  The upper and the lower
curves represent $R_o$ and $R_i$ respectively and the middle step line represents the height of the tubular condensate in unit of ribbon width $w$.}
\label{fatdisk}
\end{figure}

\begin{figure}
\onefigure[width=6cm]{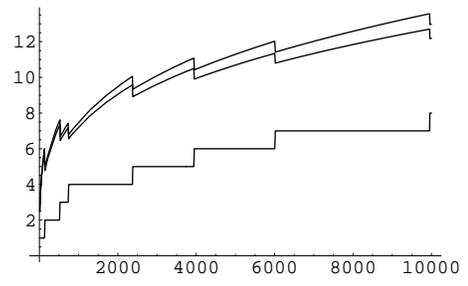}
\caption{Layering transition in a thin condition $\Gamma/\Lambda \approx 0.5$.  
The horizontal axis is
the polymer length $L/w$.   The upper two curves represent  $R_o$ and $R_i$ respectively 
and the lower step line represents the height of the tubular condensate in unit of ribbon width $w$.}
\label{thindisk}
\end{figure}

In order to show the layering transition of equilibrium form in a fat condition, 
in fig.\ref{fatdisk} we plot the outer radius of the condesates $R_o$,
the radius of the inner hole of the condensates $R_i$, 
and the quantized height of the condensate $nw$.  This fat condition is always achieved in the long chain limit.  For large $L$, the size of the condensate is roughly proportional to $\sqrt{L}$.
Below the medium chain length, there exists a certain parameter region where the condensate is
thin.  We show in fig.\ref{thindisk} behavior of the layering transition in a thin condition.%

We have assumed here that the system can form a
well-defined annulus.  It is important and interesting to examine the
fluctuations around the equilibrium shape of the system.  We will concentrate here on the case of a
single layer.   There are two  types of fluctuations. The first kind is fluctuations of the shape in the
radial direction, {\it i.e.}, distortion from a circle.   We discuss it later.  

The second kind is fluctuations out of the plane which cause two kinds of energy penalty. The first is arises from hard
bend and the energy penalty is $\epsilon_h/2\int_0^L ds (d \phi/ds)^2$, where $\phi$ is the
angle made by the tangent to the ribbon out of the plane of the
annulus. The second arises from the surface area exposed due to
differences in the height $z(s)$ of the layers. This gives rise to an
energy penalty $ \gamma_e \int_{2 \pi R_1}^L ds |z(s) - z\{s-2\pi
r(s)\}|.$ Noting that 
$z(s) = \int_0^s ds^{\prime} \phi(s^\prime) $ 
and expanding for small gradients gives $ 2 \pi \
\gamma_e \int_{2 \pi R_1}^L ds |\phi(s)| r(s). $ The total energy
penalty is then
\begin{equation}
\frac{\epsilon_h}{2} \int_0^L ds \left( d \phi \over ds \right)^2 + 2 \pi  \gamma_e \int_{2
\pi R_1}^L ds \ r(s) |\phi(s)| .
\end{equation}
This kind of total energy is peculiar
because of the form of the surface tension term, $|\phi(s)|$. This
term is first order and not differentiable about the equilibrium
position $\phi=0$. This means that the usual eigenmode techniques we
might have used to analyse the system are not applicable.

However, a good estimate of the size of the fluctuations and in
particular $\delta z \equiv \sqrt{\langle (z(L) - z(0))^2 \rangle }\ $
can be found by mapping the problem onto that of a semiflexible chain
in a nematic field.  From now on we ignore all numerical
prefactors.  For our purposes it is convenient to use an approach
developed by Odijk \cite{Odijk1, Odijk2}. He considers a wormlike chain
directed by a parabolic potential $\mu \phi^2$, where $\mu$ is a
constant. The chain has a typical angular deviation $\delta \phi
\equiv \sqrt{ \langle \phi^2 \rangle} $ . Such a chain can be viewed
as a series of steps of size $\lambda_{\parallel} = l_h (\delta
\phi)^2$ in the plane of the annulus and $\lambda_z=
\lambda_{\parallel} \delta \phi = l_h (\delta \phi)^3$ perpendicular
to the annulus.  A chain of length $L$ undergoes an anisotropic random
walk of $L/\lambda_{\parallel}$ steps in the plane, with each step
travelling a distance $\lambda_z$ perpendicular to the plane. This
implies the deviation out of the plane is $ \delta z = \sqrt{L l_h}
(\delta \phi)^2$.

Odijk shows that the strength of the potential and the angular
deviations are related by $ (\delta \phi)^4 = kT/(\mu l_h)$, so that
$\delta z = \sqrt{ kT L /\mu }$. This result cannot be applied
directly to our problem, since we have a confining potential of the
form $\gamma_e r(s) |\phi(s)|$.  However, the details of the shape of
the potential cannot matter for the gross behaviour of the system.
In particular we can replace our potential by a harmonic one, provided
we choose $\mu$ self-consistently, {\it i.e.}, we choose $\mu$ so that for a
typical value of the angular deviation the two potentials
coincide.  Equating the two potentials gives $ \gamma_e r(s) |\phi| =
\mu \phi^2$. This together, with the relation $ (\delta \phi)^4 =
kT/(\mu l_h)$ gives $\mu = (\gamma_e r(s))^{4/3} (l_h/kT)^{1/3}$. Note
that since $r$ varies along the chain, the effective confining
potential $\mu$ also varies along the chain.  We can now integrate
over the whole annulus to get the variation in height $\delta
z$. Between $r$ and $r+dr$ there is a total length $ r dr/t$ and hence
the square of the step in the $z$ direction is $ dz^2 = (kT/(\gamma_e
r))^{4/3} l_h^{-1/3} t^{-1} r dr $. Integrating over the whole annulus
gives: 
\begin{equation}
\delta z = \frac{kT}{(\gamma_e l_h)^{2/3}} \sqrt{\frac{l_h}{t}} \sqrt{
R_o^{2/3} - R_i^{2/3}}.
\end{equation}
Note that this deviation grows only very
slowly with the chain length. In, particular, for very long chains we
have $\delta z = (kT/(\gamma_e l_h))^{2/3} (l_h/t)^{1/2} t^{1/6}
L^{1/6}$.  Setting $t=a$ and $\gamma_e = g kT/a^2$ gives $ \delta z
\approx a g^{-2/3} (L/l_h)^{1/6}$. Thus, in general the overall
fluctuations will be less than, or of the same order as the annular
thickness, $w$, and the annulus will be well-defined and relatively
flat.  This is the conclusion for one layer. 

For many layers this
conclusion remains true, since in that case the effective persistence
length scales with the number of layers and the radius also scales
with the number of layers. This implies that the absolute deviation in
thickness remains roughly constant and the relative variation
decreases as more layers are added.

We now discuss fluctuations of radial direction.
The simplest mode is one where the inner radius changes from its
equilibrium value, $R_i^*$.  In the long-chain limit, where $R_i^* \ll
\sqrt{Lt}$ the energy penalty for deviating from equilibrium is 
\begin{equation}
\delta U = { \pi\epsilon_e \over 2t}  \left(  \frac{R_i - R_i^* }{\Lambda} \right)^2.
\end{equation}
 Equating this to $kT/2$ gives a typical size of
fluctuation $\sqrt{ \langle (R_i-R_i^*)^2 \rangle } = \Lambda \sqrt{t/( \pi
l_e)}$.  For $l_e/t=10$ this gives a relative fluctuation of
$13\%$.  We expect that the fluctuations within the plane will be reasonably small. 

In conclusion, ribbon polymers which have  large bending rigidities in lateral direction 
spontaneously form anisotropic condensates in poor solvents.  
The size of those peculiar condensates quantized as a variation
of the length of polymers and solvent quality.


When the material is homogeneous, the
surface tensions and elastic energies are determined by geometry
of the ribbon.  In general, for homogeneous materials, we may
assume $\epsilon_e < \epsilon_h$ and $\gamma_e \approx \gamma_h$,
which is the case we discussed in this paper.
On the chemical scale, however, other anisotropic conditions
are possible in principle.  The ribbon polymers with $\epsilon_e < \epsilon_h$ 
and $w\gamma_e < t\gamma_h$\cite{Nelson}, 
form a different type of anisotropic condensates in poor solvents,
that will be discussed somewhere else.


\acknowledgments
The authors thank Jean-Louis Sikorav for discussions and comments on the manuscript.
This work was initiated with support by NTT Basic Research Laboratories.
Y.Y.S. wishes to acknowledge CEA, IPhT and Takushoku University, RISE for financial support.


\begin{thebibliography}{99}

\bibitem{toroid0} 
\Name{Klimenko S. M., Tikkchonenko T. I. \and Andreev V. M.}
\Review{J. Mol. Biol.} \Vol{23} \Year{1967} \Page{523}

\bibitem{toroid1}
\Name{Grosberg A. Yu.}
\Review{Biophysics} \Vol{24} \Year{1979} \Page{30}

\bibitem{Frank-Kam} 
\Name{Frank-Kamenetskii M. D., Anshelevich V. V. \and Lukashin A. V.}
\Review{Sov. Phys. Usp.} \Vol{30} \Year{1987} \Page{317}

\bibitem{MarkoSiggia}
\Name{Marko J. F. \and Siggia E. D.}
\Review{Macromolecules} \Vol{27} \Year{1994} \Page{981}

\bibitem{Odijk0} 
\Name{Odijk T.} 
\Review{J. Chem. Phys.} \Vol{105} \Year{1996} \Page{1270}

\bibitem{Marko}
\Name{Marko J. F.}
\Review{Phys. Rev. E} \Vol{55} \Year{1997} \Page{1758}

\bibitem{DNAbend}
\Name{Haijun Z., Yang Z. \and Zhong-can O.-Y.}
\Review{Phys. Rev. Lett.} \Vol{82} \Year{1999} \Page{4560}

\bibitem{Exp}
\Name{Schl\"{u}ter A.-D.} 
\Review{Adv. Mater.} \Vol{3} \Year{1991} \Page{282}

\bibitem{Nyrk1}
\Name{Nyrkova I. A., Semenov A. N., Joanny J.-F. \and Khokhlov A. R.}
\Review{J. Phys. II (Paris)} \Vol{6} \Year{1996} \Page{1411}

\bibitem{Nyrk2}
\Name{Nyrkova I. A., Semenov A. N. \and Joanny J.-F.}
\Review{J. Phys. II (Paris)} \Vol{7} \Year{1997} \Page{825}

\bibitem{Nyrk3}
\Name{Nyrkova I. A., Semenov A. N. \and Joanny J.-F.}
\Review{J. Phys. II (Paris)} \Vol{7} \Year{1997} \Page{847}

\bibitem{Goles}
\Name{Golestanian R. \and Liverpool T. B.}
\Review{Phys. Rev. E} \Vol{62} \Year{2000} \Page{5488}

\bibitem{Landau}
\Name{Landau L. D. \and Lifshitz E. M.}
\Book{Theory of Elasticity}
\Publ{Pergamon, London} \Year{1970} \Page{77}

\bibitem{toroid2}
\Name{Bloomfield V. A.}
\Review{Biopolymers} \Vol{31} \Year{1991} \Page{1471}

\bibitem{toroid3}
\Name{Bloomfield V. A.}
\Review{Biopolymers} \Vol{44} \Year{1997} \Page{269}

\bibitem{Li}
\Name{Li A.-Z., Fan T-Y. \and Ding M.}
\Review{Science in China B} \Vol{35} \Year{1992} \Page{169}

\bibitem{Ubbink}
\Name{Ubbink J. \and Odijk T.}
\Review{Biophys. J.} \Vol{68} \Year{1995} \Page{54}

\bibitem{Hud}
\Name{Hud T., Downing K. H. \and Balhorn R.}
\Review{Proc. Natl. Acad. Sci. USA} \Vol{92} \Year{1995} \Page{3581}

\bibitem{Park}
\Name{Park S. Y., Harries D. \and Gelbart W. M.}
\Review{Biophys. J.} \Vol{75} \Year{1998} \Page{714}

\bibitem{toroid4}
\Name{Pereira G. G. \and Williams D. R. M.}
\Review{Biophys. J.} \Vol{80} \Year{2001} \Page{161}

\bibitem{Takenaka}
\Name{Takenaka Y., Yoshikawa K., Yoshikawa Y., Koyama Y. \and Kanbe T.}
\Review{J. Chem. Phys.} \Vol{123} \Year{2005} \Page{014902}

\bibitem{globule1}
\Name{Grosberg A. Yu. \and  Zhestkov A. V.}
\Review{J. Biomol. Struct. Dyn.} \Vol{3} \Year{1986} \Page{859}

\bibitem{globule2}
\Name{Vasilevskaya V. V., Khokhlov A. R., Kidoaki S. \and Yoshikawa K.}
\Review{Biopolymers} \Vol{41} \Year{1997} \Page{51}

\bibitem{Sikorav}
\Name{Jary, D. \and Sikorav J.-L.}
\Review{Biochem.} \Vol{38} \Year{1999} \Page{3223}

\bibitem{Odijk04}
\Name{Odijk T.}
\Review{Phil. Trans. R. Soc. Lond. A} \Vol{362} \Year{2004} \Page{1497}

\bibitem{globule3} 
\Name{Cooke I. R. \and Williams D. R. M.}
\Review{Physica A} \Vol{339} \Year{2004} \Page{45}

\bibitem{yoshikawa} 
\Name{Yoshikawa Y.,Yoshikawa K.  \and Kanbe T.}
\Review{Biophys. Chem.} \Vol{61} \Year{1996} \Page{93}

\bibitem{intermed1} 
\Name{Schnurr B., MacKintosh F. C. \and Williams D. R. M.}
\Review{Europhys. Lett.} \Vol{51} \Year{2000} \Page{279}

\bibitem{intermed2}
\Name{Pereira G. G. \and Williams D. R. M.}
\Review{ANZIAM J.} \Vol{45} \Year{2004} \Page{C163}

\bibitem{intermed3}
\Name{Yoshinaga N., Yoshikawa K. \and Ohta T.}
\Review{Euro. Phys. J. E} \Vol{17} \Year{2005} \Page{485}

\bibitem{Odijk1}
\Name{Odijk T.}
\Review{Macromolecules} \Vol{16} \Year{1983} \Page{1340}

\bibitem{Odijk2}
\Name{Odijk T.}
\Review{Macromolecules} \Vol{19} \Year{1986} \Page{2313}

\bibitem{Nelson}
\Name{Nelson J. C., Saven J. G., Moore J. S. \and Wolynes P. G.}
\Review{Science} \Vol{277} \Year{1997} \Page{1793}

\end{thebibliography}
\end{document}